\documentclass[useAMS,usenatbib]{mn2e}
\usepackage{graphicx}\usepackage{epsfig}\usepackage{epsf}
\usepackage{amsmath}\usepackage{amssymb}\usepackage{stfloats}
\usepackage{caption}
\usepackage{deluxetable}

\title[The aftermath of SN\,Hunt\,248]{The dusty aftermath of SN\,Hunt\,248: merger-burst remnant?} 
\author[Mauerhan et al.]{Jon C. Mauerhan$^{1}$\thanks{E-mail: mauerhan@astro.berkeley.edu}, Schuyler D. Van Dyk$^{2}$, Joel Johansson$^{3}$, Ori D. Fox$^{4}$, \newauthor Alexei V.\ Filippenko$^{1,5}$, Melissa L. Graham$^{6,1}$ \\
 $^{1}$Department of Astronomy, University of California, Berkeley, CA 94720-3411, USA \\
 $^{2}$Caltech/IPAC, 100-22, Pasadena, CA 91125, USA  \\
 $^{3}$Department of Particle Physics and Astrophysics, Weizmann Institute of Science, 234 Herzl St., Rehovot, Israel \\
 $^{4}$Space Telescope Science Institute, 3700 San Martin Drive, Baltimore, MD 21218, USA  \\
$^{5}$Senior Miller Fellow, Miller Institute for Basic Research in Science, University of California, Berkeley, CA 94720, USA \\
$^{6}$Department of Astronomy, University of Washington, Box 351580, Seattle, WA 98195-1580}
\begin{document}

\pagerange{\pageref{firstpage}--\pageref{lastpage}} \pubyear{2013}
\maketitle
\label{firstpage}

\begin{abstract}
SN\,Hunt\,248 was classified as a nonterminal eruption (a SN ``impostor") from a directly identified and highly variable cool hypergiant star. The 2014 outburst achieved peak luminosity equivalent to that of the historic eruption of luminous blue variable (LBV) $\eta$\,Car, and exhibited a multipeaked optical light curve that rapidly faded after $\sim100$ days. We report ultraviolet (UV) through optical observations of SN\,Hunt\,248 with the \textit{Hubble Space Telescope (HST)} about 1\,yr after the outburst, and mid-infrared observations with the \textit{Spitzer Space Telescope} before the burst and in decline. The {\it{HST}} data reveal a source that is a factor of $\sim10$ dimmer in apparent brightness than the faintest available measurement of the precursor star. The UV--optical spectral energy distribution (SED) requires a strong Balmer continuum, consistent with a hot B4--B5 photosphere attenuated by grey circumstellar extinction. Substantial mid-infrared excess of the source is consistent with thermal emission from hot dust with a mass of $\sim10^{-6}$--$10^{-5}\,{\rm M}_{\odot}$ and a geometric extent that is comparable to  the expansion radius of the ejecta from the 2014 event. SED modeling indicates that the dust consists of relatively large grains ($>0.3\,\mu$m), which could be related to the grey circumstellar extinction that we infer for the UV--optical counterpart. Revised analysis of the precursor photometry is also consistent with grey extinction by circumstellar dust, and suggests that the initial mass of the star could be twice as large as previously estimated (nearly $\sim60\,{\rm M}_{\odot}$). Reanalysis of the earlier outburst data shows that the peak luminosity and outflow velocity of the eruption are consistent with a trend exhibited by stellar merger candidates, prompting speculation that SN\,Hunt248 may also have stemmed from a massive stellar merger or common-envelope ejection. 
\end{abstract}

\section{Introduction}
``Supernova impostors" are a heterogeneous class of transient exhibiting luminosities between those of classical novae and supernovae (SNe), and a wide variety of light curves and spectral features (Smith et al. 2011; Van Dyk \& Matheson 2012). Some have been broadly characterised as extragalactic analogs to the historic super-Eddington eruptions of the Galactic luminous blue variable (LBV) stars $\eta$\,Carinae and P\,Cygni (Humphreys \& Davidson 1994; Van Dyk 2000; Smith et al. 2011; Smith et al. 2016a; Humphreys et al. 2016). However, the physical mechanisms involved remain unclear.  Indeed, the variety of transients classifiable as SN impostors suggests that there are multiple evolutionary channels. Current possibilities include instabilities associated with late-stage nuclear burning (Shiode \& Quataert 2014; Smith \& Arnett 2014), violent binary encounters (Soker 2004; Kashi \& Soker 2010; Smith \& Frew 2011), and stellar mergers involving massive binary star systems (Smith et al. 2016b; Kochanek et al. 2014; Soker \& Kashi 2013). 

Recent studies of the fading optical--infrared (IR) remnants of luminous transients have shown that objects previously classified as SN\,impostors might actually be terminal explosions after all, in which a stellar core collapses, but with an incomplete or failed expulsion of the stellar mantle. Indeed, the fate of the prototype impostor SN 1997bs (Van Dyk et al. 2000) has recently come under question, based on the unexpectedly low luminosity for the optical-IR remnant relative to the directly identified stellar precursor (Adams \& Kochanek 2015). The fate of other historic transients for which high-quality precursor data were not available have also been the source of ongoing debate (e.g., SN\,1961V; Smith et al. 2011; Kochanek et al. 2011; Van Dyk \& Matheson 2012b), in part because the cooling outflows from nonterminal eruptions can form dust that obscures the star, and also because late-time line emission can result from persistent interaction between the outflow and an extended distribution of slower pre-existing circumstellar material (CSM). These issues underscore the importance of obtaining late-time multiwavelength monitoring observations of SN impostors, in order to track their post-outburst evolution and determine their ultimate fate.

\begin{figure}
\includegraphics[width=3.3in]{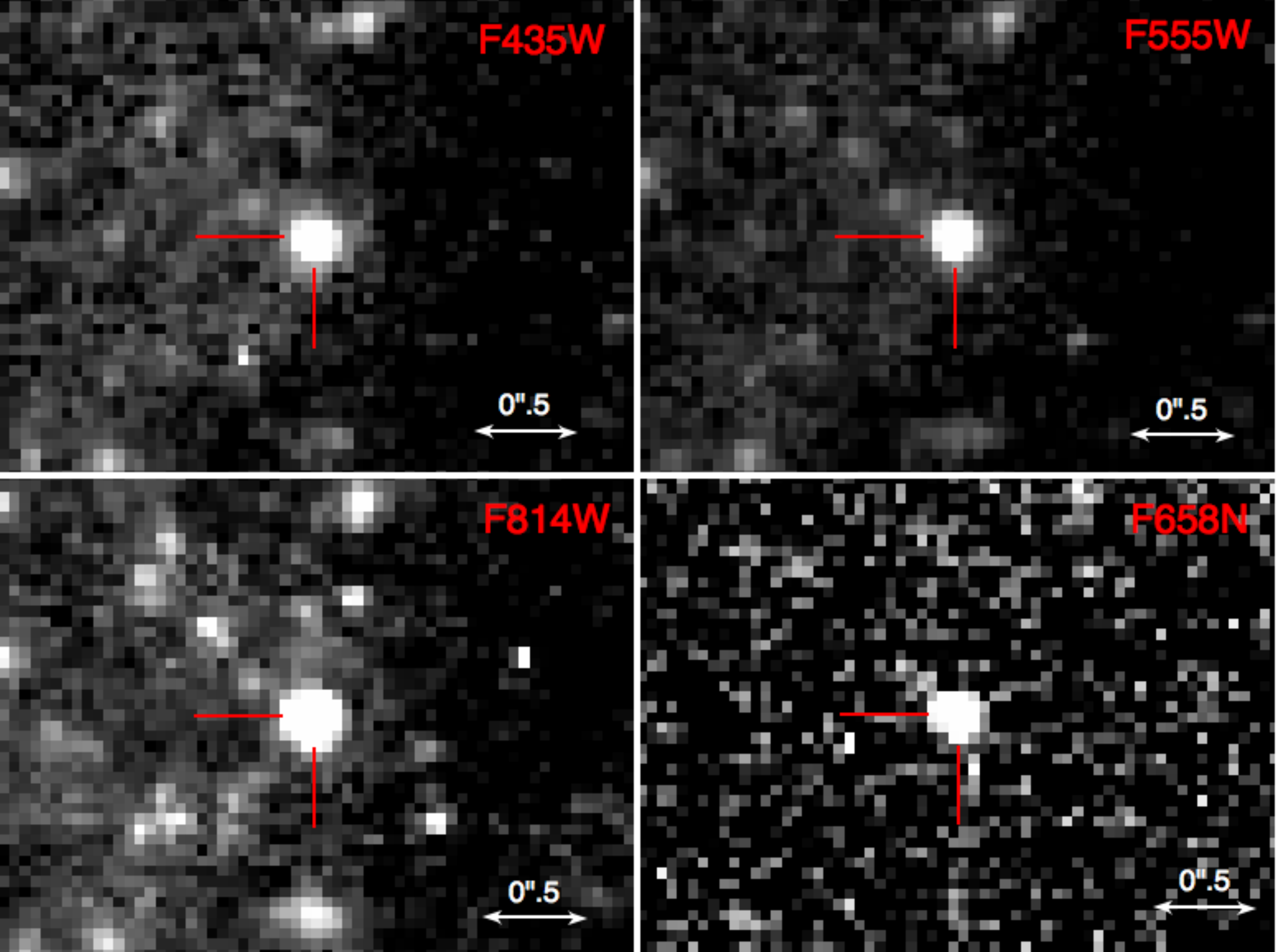}
\caption{{\it HST}/WFPC2 and ACS images (log stretch) of the cool hypergiant precursor of SN\,Hunt\,248 (reproduced from data presented by Mauerhan et al. 2015), from 3374 days before the onset of the 2014 eruption (broad-band filter images; the F658N image is from 3715 days before eruption).  North is up and east is toward the left. }
\label{fig:precursor}
\end{figure}

\begin{figure}
\includegraphics[width=3.3in]{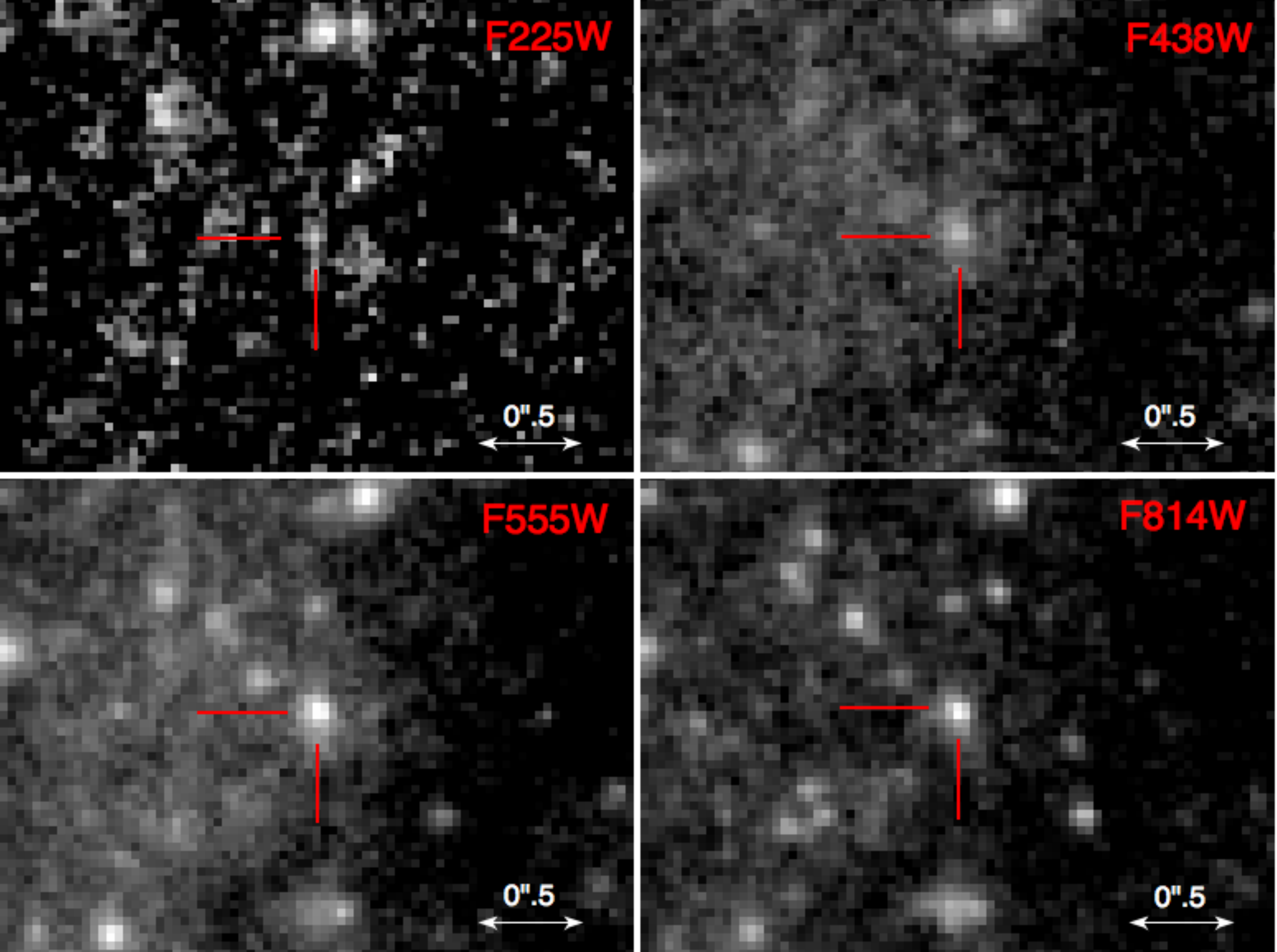}
\caption{{\it HST}/WFC3 images (log stretch) of the remnant of SN\,Hunt\,248, $\sim1$\,yr after the peak of the 2014 eruption. North is up and east is toward the left.}
\label{fig:remnant}
\end{figure}

\begin{center}
 \begin{table} 
      \caption{{\it HST} photometry of the remnant of SN\,Hunt\,248.}
\renewcommand\tabcolsep{3.5pt} \scriptsize
\begin{tabular}[b]{@{}lcccc}
\hline
\hline
Instrument/Band & Magnitude & Flux ($\mu$Jy) & MJD & Epoch (days) \\
\hline
\hline
WFC3/F225W   &$25.16\pm0.09$ & $0.068\pm 0.006 $  & 57204.05 & 374 \\  
WFC3/F438W  & $25.84\pm0.05$   & $0.193\pm 0.010 $  &57204.00 & 374 \\
WFC3/F555W  & $25.46\pm0.03$   &  $0.243\pm 0.007 $ &57199.87 & 370 \\
ACS/F814W  & $24.51\pm0.04$ &  $0.386\pm 0.015 $ &57200.38  & 371 \\  
\hline
\end{tabular}\label{tab:p48} 
\begin{flushleft}
 \scriptsize$^\textrm{a}$Uncertainties are statistical. Epochs are given as days from $V$-band peak (MJD\,56830.3; Mauerhan et al. 2015). \\
\end{flushleft}
\end{table}
\end{center}

SN\,Hunt\,248 was a luminous transient in NGC\,5806 classified as a SN impostor. The light curve exhibited a main peak equivalent in luminosity to the peak of $\eta$\,Car's historic outburst in the 1840s, and another subsequent peak of longer duration that was likely the result of interaction between the erupted material and slower pre-existing CSM expelled prior to the outburst (Mauerhan et al. 2015; Kankare et al. 2015). A particularly interesting aspect of SN\,Hunt\,248 is the detection of the luminous precursor star in archival data, shown in Figure\,\ref{fig:precursor} (images reproduced from Mauerhan  et al. 2015). Multicolour photometry from the {\it Hubble Space Telescope (HST)} showed that the stellar precursor's position on the Hertzsprung-Russell (HR) diagram was consistent with that of a cool hypergiant star. The subsequent giant eruption from the star in 2014 provided observational support for a hypothesis that cool hypergiants might actually be relatively hot LBV stars enshrouded in an opaque wind that creates an extended pseudophotosphere (Smith \& Vink 2004). Detailed study of the aftermath of the eruption thus provides an interesting opportunity to probe the post-outburst state and recovery of the stellar remnant. 

Here we present ultraviolet (UV) through IR observations of SN\,Hunt\,248 with the \textit{HST} and the \textit{Spitzer Space Telescope} about 1\,yr after the giant outburst. In \S3 we model the mid-IR data as a source of thermal dust emission. In \S4 we discuss the effects of circumstellar extinction and implications for the nature of the remnant star. The times of all observation epochs are presented as days past $V$-band peak on 2014 June 21 (MJD 56830.3; UT dates are used throughout this paper). A foreground interstellar extinction value of $A_V=0.14$\,mag has been adopted (Mauerhan et al. 2015).

\section{Observations}

\subsection{{\it HST} Imaging}
High-resolution imaging observations of SN\,Hunt\,248 were performed with the {\it HST} Wide-field Camera 3 ({\it HST}/WFC3) on 2015\,June\,26 and 30 (369 and 374 days after the peak of the 2014 eruption) under {\it HST} programmes GO-13684 and GO-13822 (PIs S. Van Dyk and G. Folatelli, respectively).  Exposures were obtained in the F225W (NUV), F438W ($B$), F555W ($V$), and F814W ($I$) filters. A point source at the position of SN\,Hunt\,248 is securely detected in all bands, as shown in Figure\,\ref{fig:remnant}. Photometry of the source was extracted from the images using {\sc{dolphot}} (Dolphin 2000). We tried two different approaches to estimate the background, including the use of an annulus region to measure the sky (FitSky=1) and, alternatively, measuring the sky within the point-spread function (PSF) aperture (FitSky=3, best to use when the field is very crowded). Our annulus-based background subtraction produced the most consistent results for all bands, although the results from each setting are within the respective uncertainties. The photometry is listed in Table\,1. 

\begin{figure*}
\includegraphics[width=7in]{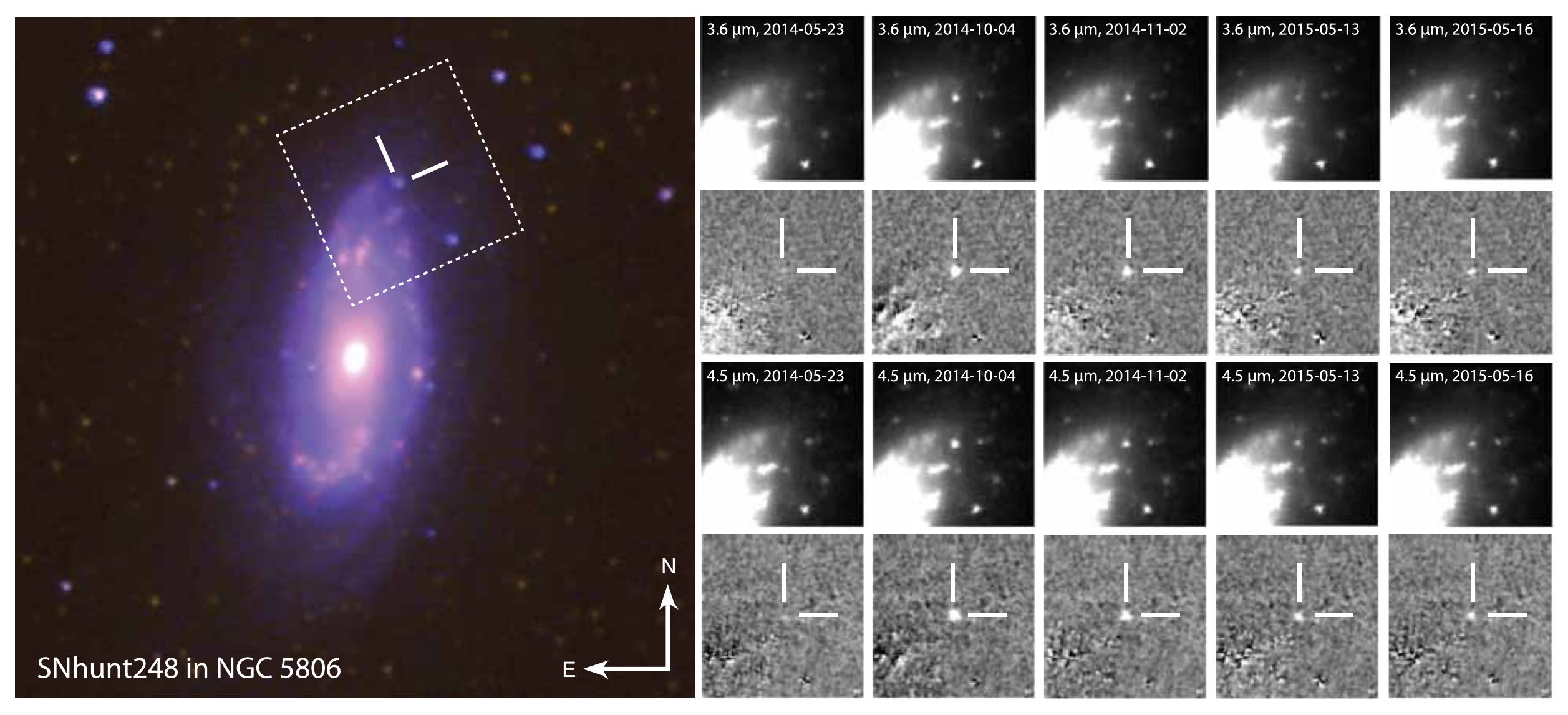}
\caption{Colour composite of the 3.6\,$\mu$m (green) and 4.5\,$\mu$m (red) \textit{Spitzer}/IRAC template images of NGC\,5806 and a Palomar Transient Factory $R$-band image (blue), and the template-subtracted images of the region around SN\,Hunt\,248 (tiled frames).}
\label{fig:spitzer}
\end{figure*}

 \begin{center}
 \begin{table} 
      \caption{\textit{Spitzer}/IRAC photometry of SN\,Hunt\,248.}
\renewcommand\tabcolsep{4.pt} \scriptsize
\begin{tabular}[b]{@{}lrrrl}
\hline
\hline
MJD & Epoch (days) & 3.6\,$\mu$m & 4.5\,$\mu$m & Programme ID (PI) \\
\hline 
\hline
55066.9 & $-1763$	&	 $<9.49$    &  $<5.99$ & 61063 (Sheth)\\  
56800.7 & $-30$	&	 $19.35\pm6.11$    &  $11.81\pm5.09$& 10152 (Kasliwal) \\  
56934.6 & 104	&	$110.26\pm6.38$   &	$96.41\pm5.86$ & 10152 (Kasliwal) \\ 
56963.4 & 133      &	 $65.56\pm6.14$    &	$59.70\pm5.91$ & 10139 (Fox) \\ %
57155.5 & 325	 &	 $32.72\pm7.82$     &$29.26\pm4.76$ & 11053 (Fox)\\ %
57158.2 & 328	& 	 $34.30\pm5.44$     &$29.57\pm5.61$ & 11053 (Fox) \\ %
\hline
\end{tabular}\label{tab:p48} 
\begin{flushleft}
 \scriptsize$^\textrm{a}$Fluxes are in units of $\mu$Jy. Uncertainties are statistical. Epochs are given as days from $V$-band peak (MJD\,56830.3; Mauerhan et al. 2015). \\
\end{flushleft}
\end{table}
\end{center}

\subsection{\textit{Spitzer} Imaging}
SN\,Hunt\,248 was observed on five epochs during the \textit{Spitzer Space Telescope} Warm Mission utilising channels 1 (3.6\,$\mu$m) and 2 (4.5\,$\mu$m) of the Infrared Array Camera (IRAC; Fazio et al. 2004). We acquired fully coadded and calibrated data from the {\it Spitzer} Heritage Archive\footnote{http://sha.ipac.caltech.edu/applications/Spitzer/SHA/} from programme IDs 61063 (PI K. Sheth), 10152 (PI M. Kasliwal), and 11053 (PI O. Fox). The images for all epochs were registered with an earlier pre-outburst image of the host galaxy, which was used as a subtraction template. The template-subtracted images are shown in Figure\,\ref{fig:spitzer}. We performed aperture photometry on the template-subtracted (PBCD / Level 2) images using a 6-pixel aperture radius and aperture corrections listed in Table\,4.7 of the \textit{Spitzer} IRAC Instrument Handbook\footnote{http://irsa.ipac.caltech.edu/data/SPITZER/docs/irac/}. The infrared photometry are listed in Table~2.

\begin{figure*}
\includegraphics[width=5.7in]{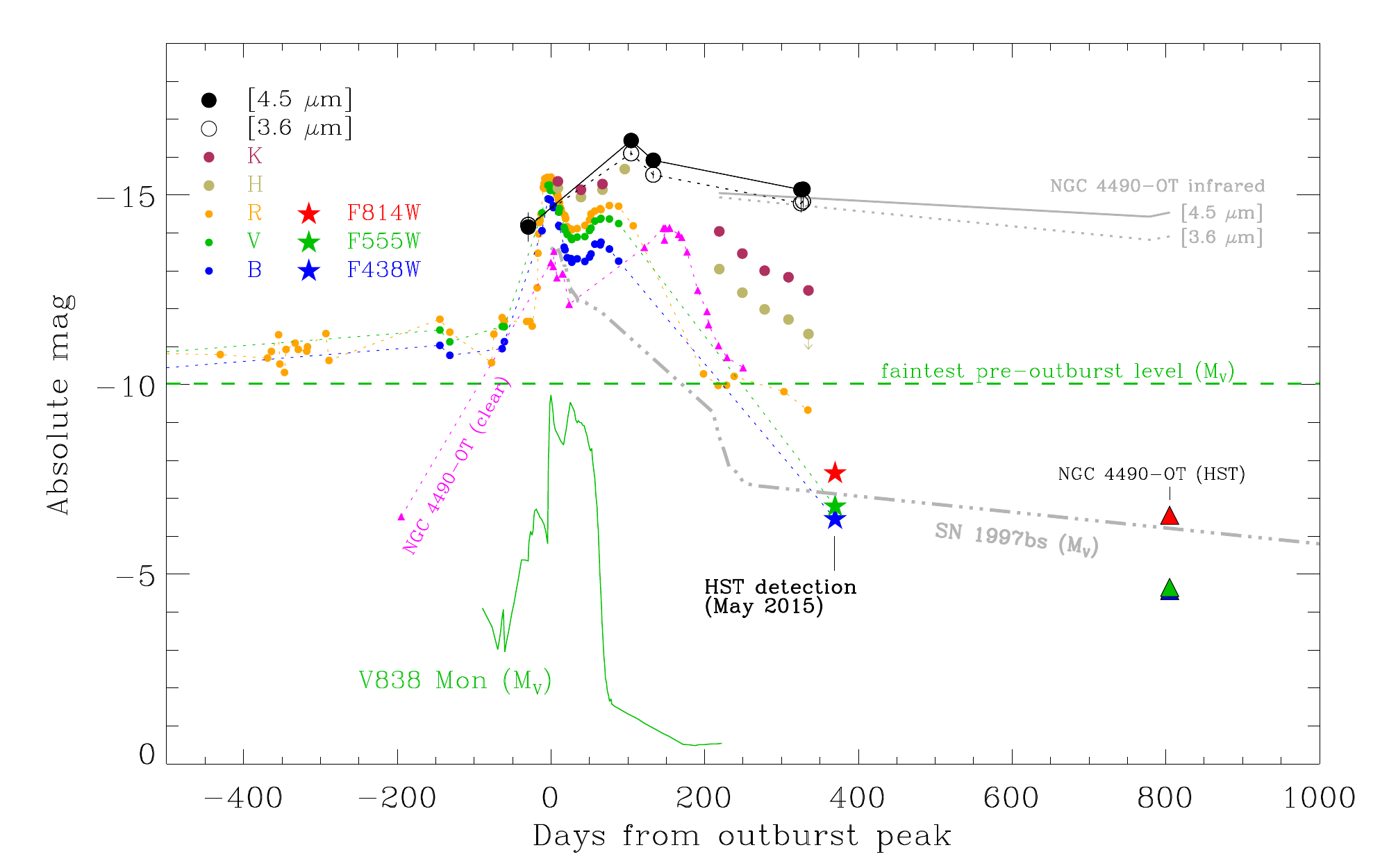}
\caption{Long-term light curve of SN\,Hunt\,248 (coloured filled circles), including the late-time mid-IR \textit{Spitzer} (black open and filled circles) and UV--optical {\it HST} data (coloured 5-pointed stars) presented here. Optical and near-IR photometric data of the precursor and main outburst are from Mauerhan et al. (2015) and Kankare et al. (2015). The optical photometry of NGC\,4490-OT is also shown as coloured triangles (magenta is ground-based clear-filter photometry; red, green, and blue are (respectively) late-time F814W, F555W, and F438W filter photometry from {\it HST}; see Smith et al. 2016b); mid-IR \textit{Spitzer} data on NGC\,4490-OT are shown as grey solid and dotted curves. The optical light curve of SN\,1997bs is displayed for comparison, including SN\,1997bs (dashed triple-dotted grey curve; Van Dyk et al. 2000). The green horizontal dashed line represents the faintest pre-outburst $V$-band absolute magnitude of the precursor star (see Mauerhan et al. 2015). The $V$-band light curve of the purported stellar merger V838\,Mon is also shown (green solid curve; Bond et al. 2003).}
\label{fig:lc}
\end{figure*}

\begin{figure}
\includegraphics[width=3.2in]{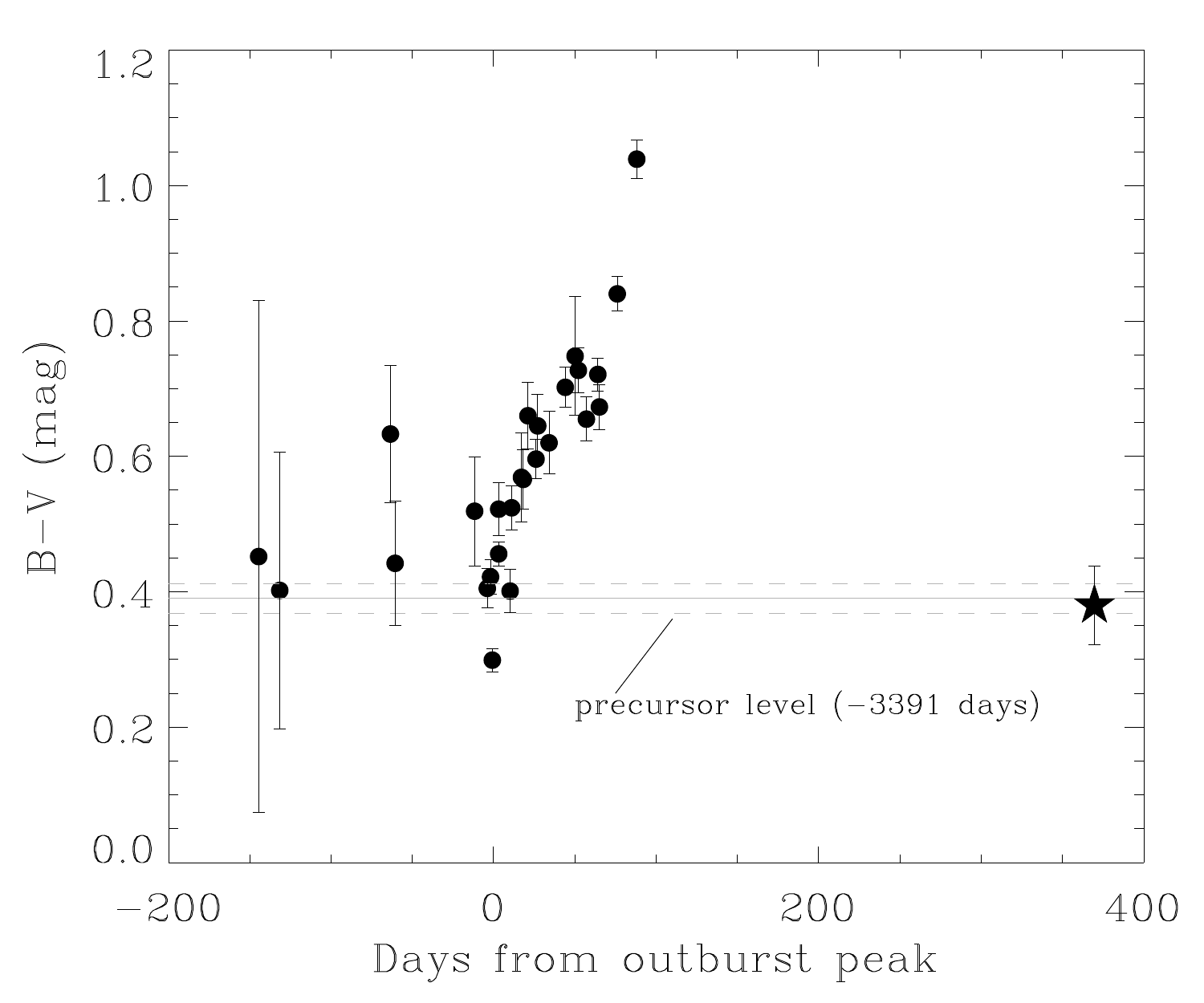}
\caption{$B-V$ colour evolution of SN\,Hunt\,248. The horizontal lines represent the value (thick line) and uncertainty envelope (thinner lines) of the stellar precursor detected with {\it HST} at $-$3391 days (see Mauerhan et al. 2015). Filled dots are data from Kankare et al. (2015). The filled 5-pointed star symbol represents our most recent measurement from Table\,1, which exhibits the same value as the stellar precursor.}
\label{fig:color}
\end{figure}

\begin{figure}
\includegraphics[width=2.9in]{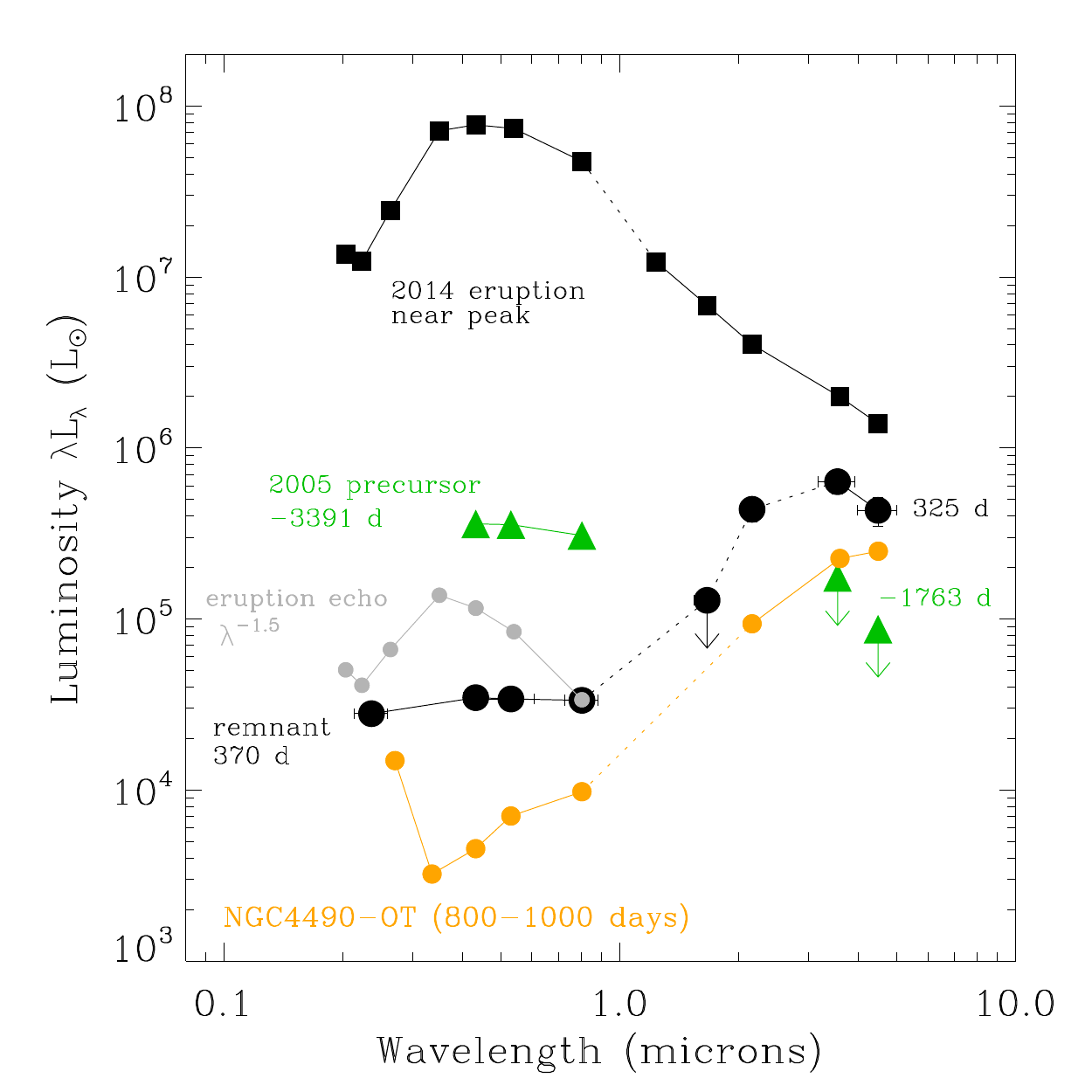}
\caption{UV--IR SED of SN\,Hunt\,248, including the 2014 outburst (black squares), the precursor (green triangles), and the remnant (black circles). The expected UV--optical SED of an echo of the 2014 eruption (see text) is shown as grey circles. The SED of massive stellar merger candidate NGC 4490-OT is also shown for comparison (orange filled circles; data from Smith et al. 2016, and reddened by adopting their extinction estimate with the extinction relation of Cardelli, Clayton, \& Mathis (1989)). }
\label{fig:sed}
\end{figure}

\section{Results \& Analysis}
The absolute-magnitude light curve of SN\,Hunt\,248 is shown in Figure\,\ref{fig:lc}, including data from Mauerhan et al. (2015) and Kankare et al. (2015).  At $\sim1$\,yr after the peak of the 2014 outburst, the source has dropped to a brightness of $V=25.46\pm0.03$\,mag, which is a factor of $\sim10$ fainter in the optical than the faintest pre-outburst state ever measured for the stellar precursor in the year 2005 ($V=22.91\pm0.01$\,mag; see Mauerhan et al. 2015). Yet, as illustrated in Figure\,\ref{fig:color}, the $B-V$ colour of $0.38\pm0.06$\,mag is consistent with no change from the precursor value of $0.39\pm0.02$\,mag, while the $V-I$ colour of $0.95\pm0.05$\,mag has become only slightly redder from the precursor value of $0.81\pm0.01$\,mag.

The latest epoch of near-IR $H$ and $K$ photometry from Kankare et al. (2015) nearly coincides with our {\it HST} UV--optical data from days 369--374 and \textit{Spitzer} mid-IR photometry from 325. We thus combined these data to construct a spectral energy distribution (SED) for the source, shown in Figure\,\ref{fig:sed}; the strong IR component of the SED is clearly seen.

\begin{table*} 
    \caption{Dust-model parameters for SN\,Hunt\,248 at epochs 104--328 days post-peak$^a$.}
\tabletypesize{\footnotesize}
\begin{tabular}[b]{@{}lccccccc} 
  & \multicolumn{3}{c}{Graphite} & & \multicolumn{3}{c}{Silicates} \\
  & \multicolumn{3}{c}{------------------------------------------} & & \multicolumn{3}{c}{------------------------------------------} \\
{$a$ ($\mu$m)} & {$T_{\rm d}$ (K)} & {$M_{\rm d}$ (M$_{\odot}$)} & {$L_{\rm d}$ (L$_{\odot}$)} & &{$T_{\rm d}$ (K)} & {$M_{\rm d}$ (M$_{\odot}$)} & {$L_{\rm d}$ (L$_{\odot}$)} \\
\hline
\multicolumn{8}{c}{104 days}\\
\hline
0.10 & 868   & 2.7e-5 & 3.0e+6  & & 1247 & 4.1e-5 & 5.5e+6   \\
0.30 & 865   & 9.4e-6 & 2.6e+6  & & 1140 & 4.9e-5 & 5.5e+6   \\
0.50 & 1086 & 3.6e-6 & 2.6e+6  & & 1038 & 5.6e-5 & 5.2e+6   \\
0.75 & 1636 & 1.6e-6 & 4.7e+6  & & 940 & 6.4e-5 & 4.5e+6   \\
1.00 & 1670 & 5.9e-7 & 5.2e+6  & & 856 & 7.3e-5 & 3.7e+6   \\

\hline
\multicolumn{8}{c}{133 days}\\
\hline
0.10 & 830  & 2.0e-5 & 1.7e+6 & & 1071 & 3.1e-5 & 3.2e+6  \\
0.30 & 827  & 7.0e-6 & 1.5e+6  & & 981 & 3.7e-5 & 3.2e+6  \\
0.50 & 1024 & 2.7e-6 & 1.5e+6  & & 894 & 4.2e-5 & 3.0e+6  \\
0.75 & 1487 & 1.2e-6 & 2.4e+6  & & 819 & 4.8e-5 & 2.7e+6  \\
1.00 & 1514 & 1.6e-6 & 2.6e+6  & & 870 & 5.4e-5 & 2.3e+6  \\
\hline
\multicolumn{8}{c}{325 days}\\
\hline
0.10 & 846 & 9.0e-6 & 8.8e+5 & &1199 & 1.4e-5 & 1.6e+6    \\
0.30 & 843 & 3.2e-6 & 7.5e+5  & &1101 & 1.7e-5 & 1.6e+6   \\
0.50 & 1050 & 1.2e-6 & 7.6e+5  &  & 1006 & 1.9e-5 & 1.5e+6  \\
0.75 & 1549 & 5.4e-7 & 1.3e+6  & & 914 & 2.2e-5 & 1.3e+6   \\
1.00 & 1579 & 7.2e-7 & 1.4e+6  & & 835 & 2.5e-5 & 1.1e+6   \\
\hline
\multicolumn{8}{c}{328 days}\\
\hline
0.10 & 882 & 7.7e-6 & 9.7e+5   & &1279 & 1.2e-5 & 1.7e+6   \\
0.30 & 879 & 2.7e-6 & 8.2e+5 & &1166 & 1.4e-5 & 1.7e+6    \\
0.50 & 1109 & 1.0e-6 & 8.3e+5  & & 1059 & 1.6e-5 & 1.6e+6  \\
0.75 & 1697 & 4.4e-7 & 1.6e+6   & & 957 & 1.9e-5 & 1.4e+6   \\
1.00 & 1733 & 5.9e-7 & 1.7e+6   & & 870 & 2.1e-5 & 1.2e+6   \\
\hline
\end{tabular}

\begin{flushleft}
 \scriptsize$^\textrm{a}${Only upper limits on dust parameters were obtainable for our earliest epoch at $-30$ days post-peak (not listed; see text). Average uncertainties for fit parameters $M_{\rm d}$, $T_{\rm d}$, and $L_{\rm d}$ are estimated at 30\%, 25\%, and 30\%, respectively (see text).}
\end{flushleft}
\end{table*}   

\begin{figure*}
\includegraphics[width=3.3in]{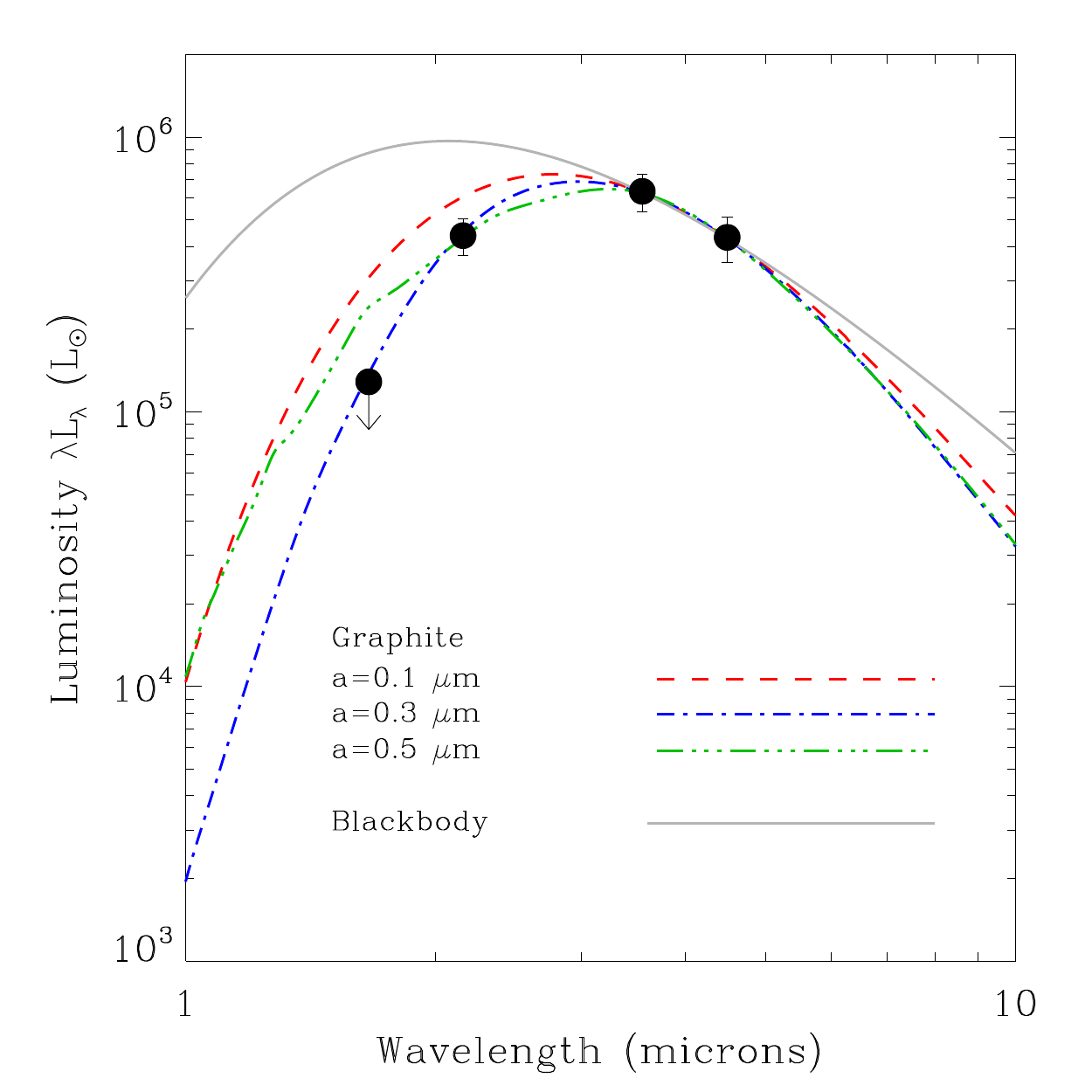}
\includegraphics[width=3.3in]{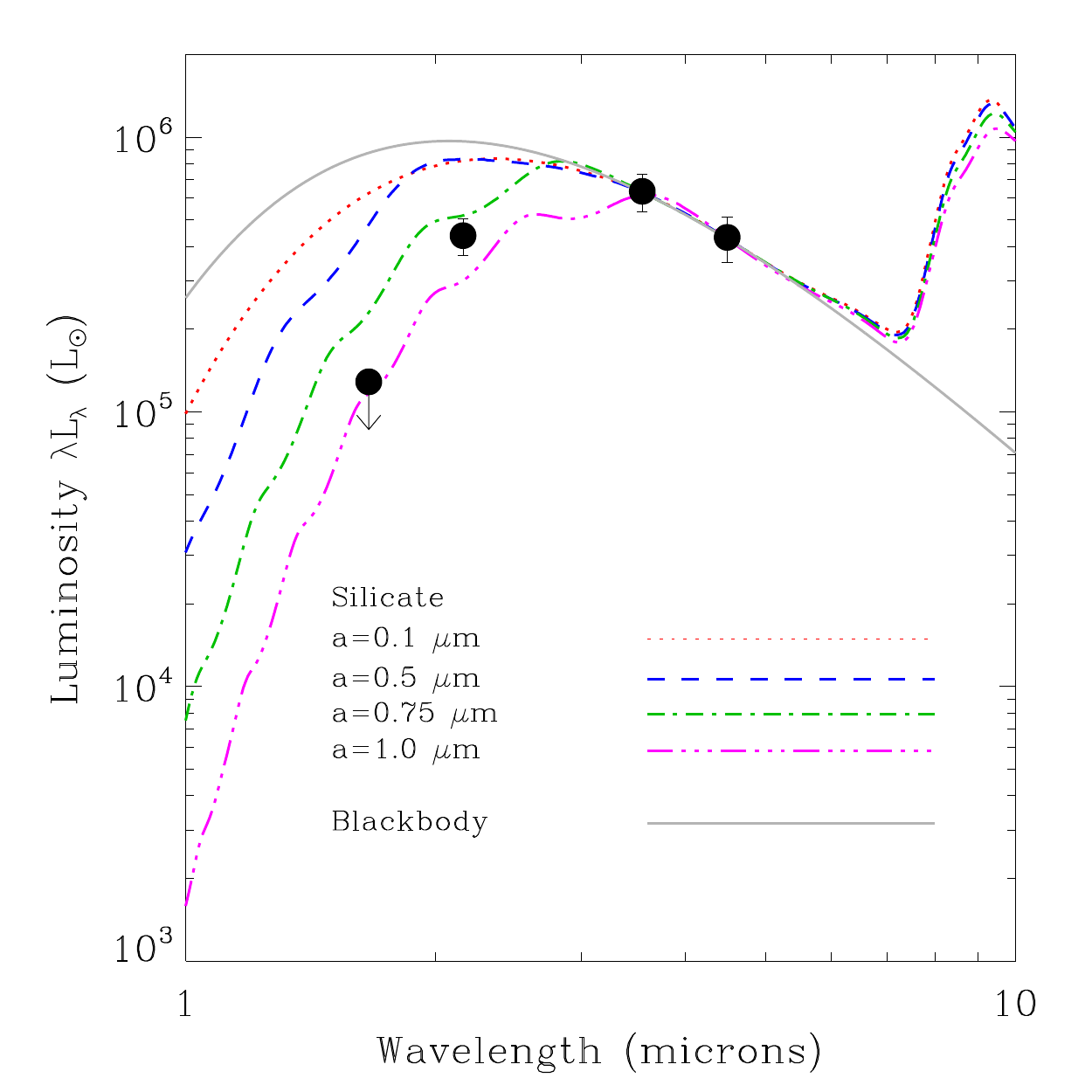}
\caption{Infrared photometry of SN\,Hunt248 on day 328 post-peak compared with the SEDs of graphite (left panel) and silicate (right panel) model dust sources. $H$ and $K$ points are day-332 measurements from Kankare et al. (2015). }
\label{fig:modfit}
\end{figure*}

\subsection{Dust modeling}
The \textit{Spitzer} data were analysed under the assumption that the source of the mid-IR emission is hot dust. We fit the SED using simple models for graphite and silicate composition (Fox et al. 2010, 2011),
with dust mass ($M_{\rm d}$) and temperature ($T_{\rm d}$) as free parameters. The flux is given by

\begin{equation}
\label{eqn:flux2}
F_\nu = \frac{M_{\rm d}\,B_\nu(T_{\rm d})\,\kappa_\nu (a)} {d^2},
\end{equation}

\noindent where $\kappa_\nu (a)$ is the dust absorption coefficient as a function of grain radius, and $d$ is the distance of the dust from the observer (Hildebrand 1983).  We performed our calculations for grain sizes in the range 0.1--1.0\,$\mu$m, looking up their associated $\kappa_\nu (a)$ values from the Mie scattering derivations discussed by Fox et al. (2010, see their Figure\,4). For simplicity, we assume optically-thin dust of a constant grain radius and emitting at a single equilibrium temperature (e.g., Hildebrand 1983). The data were fit using the IDL routine {\tt MPFIT}. Table~3 lists the best-fitting parameters for $T_{\rm d}$, $M_{\rm d}$, and the dust luminosity $L_{\rm d}$ for graphite and silicates over a range of grain radii, for epochs 104 days through 328 days. The average statistical uncertainties for $M_{\rm d}$, $T_{\rm d}$, and $L_{\rm d}$ are estimated at $\sim30$\%, $\sim25$\%, and $\sim30$\%, respectively. This estimate was obtained by performing several fits on the \textit{Spitzer} data after offsetting the photometry by the photometric errors.

For our earliest epoch just before the onset of the main eruption, 30 days before peak, satisfactory model fits for $M_{\rm d}$ and $T_{\rm d}$ were not obtainable, limiting the luminosity to $L<2\times10^6\,{\rm L}_{\odot}$. For the successfully modeled epochs thereafter, we measure no significant change with time for the dust parameters of a given model, within our quoted uncertainty ranges. For the models of graphite dust grains with $a=0.1$ and 0.3\,$\mu$m, the temperature remains 800--900~K, and inferred dust masses range between $\sim3\times10^{-6}\,{\rm M}_{\odot}$ and $\sim3\times10^{-5}\,{\rm M}_{\odot}$, with luminosities on the order of a few $\times10^6\,{\rm M}_{\odot}$. For larger grain sizes of $a=0.5$\,$\mu$m up to 1\,$\mu$m, the range of potential temperatures is hotter (1024--1720\,K). The masses of these larger grain models are systematically lower by a factor of a few, while the luminosities are comparable to those of the smaller grain models. For silicate grains, the model masses, temperatures, and luminosities are all slightly higher than for graphite---most notably for dust mass. However, the temperatures of the larger grain silicates are comparable to those of the smaller grain graphite models. We note, however, that $M_{\rm d}$ should probably be regarded as a lower limit, since there might also be a cooler component of dust to which our \textit{Spitzer} observations at 3.6\,{$\mu$m} and 4.5\,{$\mu$m} are not sensitive. 

Although our model parameters were fit using only the \textit{Spitzer} photometry at 3.6 and 4.5~$\mu$m, the epoch on day 328 post-peak was only 5 days before a ground-based near-IR $H$ and $K$ measurement from Kankare et al. (2015), so we used those data to further discriminate between the various SED models. This last epoch is also particularly important in that it is close in time to our UV--optical {\it HST} photometry of the remnant, and so can be used to estimate the expected UV--optical extinction from the dust parameters we derived (see \S4.1.2). As shown in Figure\,\ref{fig:modfit}, the results suggest that the average grain size for both the silicate and graphite models is likely to be substantially larger than the  0.1\,$\mu$m average grains size. Indeed, simple blackbody distributions of any temperature are too broad to fit the SED of SN\,Hunt248, and dust models for small grain sizes are also too broad and significantly overestimate the flux in the near-IR; the source is clearly a greybody. For graphite, the SED appears most consistent with 0.3\,$\mu$m grains, while for silicate dust, even larger grain sizes in the range 0.75--1.0\,${\mu}$m appear to provide the best match to the day 328 data.

The size of the emitting region can be estimated by considering the radius of an equivalent blackbody having luminosity and temperature indicated by the model fits,
\begin{equation}
r_{\rm bb} = \bigg(\frac{L_{\rm d}}{4 {\rm \pi} \sigma T_{\rm d}^4}\bigg)^{1/2}.
\end{equation}

\noindent Focusing on the last epoch at 328 days, the best-matching graphite ($a=0.3\,\mu$m) and silicate ($a=0.75$--1.0\,$\mu$m) dust models indicate respective radii of $2.7\times10^{15}$\,cm and (3.1--3.3) $\times10^{15}$\,cm. We therefore assume an approximate value of $3\times10^{15}$\,cm for the following analysis and interpretation.

\section{Discussion}
\subsection{The nature of the remnant}

\subsubsection{The origin of the dust}

The SED of the UV--IR remnant of SN\,Hunt\,248, shown in Figure\,\ref{fig:sed}, appears very similar to that of NGC\,4490-OT (Smith et al. 2016b). In both cases, the dust is hot and emits like a greybody, and the UV--optical counterpart  has noteworthy UV flux, especially NGC 4490-OT.  As shown in Figures~\ref{fig:lc} and \ref{fig:sed}, the mid-IR brightness evolution of both objects exhibits very similar plateaus, and they both have IR luminosities that are comparable to the optical luminosities of their directly identified stellar precursors; taken at face value, this appears consistent with heating of the dust by a luminous surviving star. Interestingly, the estimated radius of the dust ($3\times10^{15}$\,cm) matches the expected expansion radius of the ejecta after 328 days, considering the measured outflow speed of $v\approx1200$\,km\,s$^{-1}$ (Mauerhan et al. 2015). Thus, the measurements seem consistent with dust condensation in the ejecta from the 2014 event. 

\begin{figure}
\includegraphics[width=3.5in]{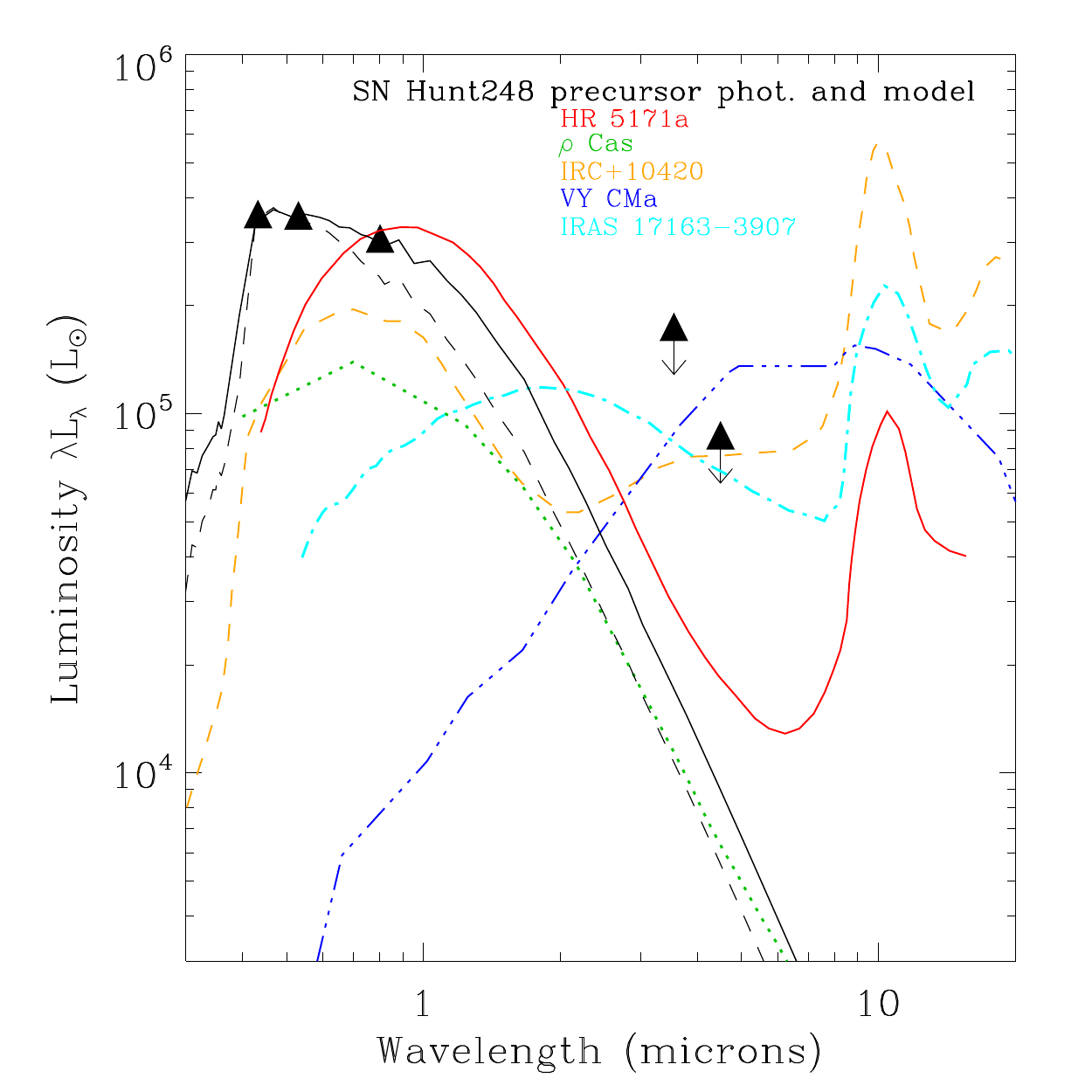}
\caption{Optical luminosity of the stellar precursor of SN\,Hunt\,248 (filled triangles; Mauerhan et al. 2015) and its mid-IR limits (black filled triangles with downward-facing arrows). The photometry was corrected only for interstellar extinction ($A_V=0.14$\,mag, $R_V=3.1$). The solid black curve is the SED of a star with $T_{\rm eff}=7000$\,K (Castelli \& Kurucz 2003), scaled to $L/{\rm L}_{\odot}=6.07$, and reddened by an additional component of grey circumstellar extinction ($A_V=0.86$\,mag, $R_V=5.4$). The SED equivalent to the previously estimated stellar parameters from Mauerhan et al. (2015), which did not account for circumstellar extinction, is represented by the black dashed curve. For comparison purposes, we also show the SEDs for the Galactic cool-warm hypergiants VY\,CMa, $\rho$\,Cas, and IRC$+$10420 (blue dashed triple-dotted, green dotted, and orange dashed curves, respectively; Shenoy et al. 2016),  HR5171a (red solid curve; Humphreys et al. 1971), and IRAS\,17163$-$3907 (cyan dashed-dotted curve; Lagadec et al. 2011). The following distances were used to calculate the luminosity: SN\,Hunt\,248 (26.4\,Mpc; Mauerhan et al. 2015), $\rho$\,Cas (2.5\,kpc; Humphreys 1978); VY\,CMa (1.2\,kpc; Shenoy et al. 2016), IRC$+$10420 (5\,kpc; Shenoy et al. 2016), HR5171 (3.6\,kpc; Chesneau et al. 2014a), and IRAS\,17163$-$3907 (4.2\,kpc, average of range estimate from Lagadec et al. 2011). The IRAS\,17163$-$3907 data were corrected for interstellar extinction in this work, adopting $A_V=2.1$\,mag (Lagadec et al. 2011) and the extinction relation of Cardelli, Clayton, \& Mathis (1989). The other SEDs from the literature account only for interstellar extinction.}
\label{fig:pre_sed}
\end{figure}

Alternatively, pre-existing dust may have been swept to large radius by the ejecta. After all, the spectra near peak brightness did exhibit the signatures of CSM interaction (see Mauerhan et al. 2015). Assuming a gas-to-dust ratio of 100, the dust masses inferred by our model ($\sim10^{-6}$ to $10^{-5}\,{\rm M}_{\odot}$) imply a total CSM mass of $\sim10^{-4}$ to $10^{-3}\,{\rm M}_{\odot}$. Therefore, if the 2014 eruption ejected only 0.1\,M$_{\odot}$ (which would be modest compared to the $>10\,{\rm M}_{\odot}$ ejected by $\eta$\,Car's historic event), the pre-existing CSM would not be massive enough to effectively decelerate the ejecta. Dust in the circumstellar environment could therefore have been swept to the expansion radius, if the grains survived the UV radiation and shock of the event. We speculate that this might explain the relatively large sizes of dust grains inferred by our models for the IR emission---i.e., the smallest circumstellar grains could have been destroyed by the 2014 outburst, leaving a distribution skewed toward larger sizes.

Unfortunately, our \textit{Spitzer} mid-IR upper limits on the stellar precursor are not sufficiently deep to provide a meaningful constraint on the pre-existing dust mass, so we cannot tell if the dust mass was lower before the eruption than in the aftermath. For example, Figure\,\ref{fig:pre_sed} shows the SED of the precursor to SN\,Hunt\,248 along with those of several Galactic cool hypergiants that have measured circumstellar dust masses in the literature. Our limits are mutually consistent with a system like $\rho$\,Cas, which has a rather low estimated dust mass of $\sim3\times10^{-8}$\,M$_{\odot}$ (Jura \& Kleinmann 1990), and more extreme dusty systems like IRC$+$10420 (Shenoy et al. 2016) and IRAS\,17163$-$3907 (Lagadec et al. 2011), the latter of which has a much larger dust mass of $\sim0.04$\,M$_{\odot}$. The comparison in Figure\,\ref{fig:pre_sed} does, however, suggest that the IR excess from a system such as VY\,CMa, with a total dust mass of $\sim0.02$\,M$_{\odot}$ (Harwit et al. 2001; Muller et al. 2007), would have been detectable at 4.5\,${\mu}$m. It is therefore plausible that the $\sim10^{-5}$\,M$_{\odot}$ dust mass we inferred {\it post} eruption could have been pre-existing, yet not detectable by our \textit{Spitzer} observations. 

Finally, we should address the possibility that the IR (and perhaps optical) emission of the remnant is the result of a light echo of the 2014 outburst off of outer dusty CSM. In such a scenario there is both delayed scattering of UV--optical light and thermal IR reprocessing of the fraction of light that gets absorbed by the dust. However, assuming that such an echo is dominated by light from the peak of the outburst and obeys a $\propto\lambda^{-1.5}$ wavelength dependence (e.g., see Fox et al. 2015), while suffering the same extinction as the precursor, the expected UV--optical SED is totally inconsistent with the observed SED at $+$370 days (see Figure\,\ref{fig:sed}, grey curve). The thermal-IR remnant also appears to be inconsistent with thermal reprocessing of an echo, as the dust temperature requires a luminosity that is far above the peak of the 2014 event. This was determined using the same line of reasoning invoked for the analysis of the remnant of NGC\,4490-OT (see Smith et al. 2016, their \S3.2.3). Assuming the ratio of the efficiencies of UV absorption to IR emission is $Q_{\textrm{UV}}/Q_{\textrm{IR}}=0.3$ (Smith et al. 2016b), the luminosity required to heat dust at a distance $r$ to a temperature $T$ can be expressed as $L/{\rm L}_{\odot}\approx{5.7\times10^{12}}\,(T_d/1000\,\textrm{K})^4\,(r/\textrm{pc})^2$. At 328 days the minimum distance of the echo-heated dust is $r\approx0.3$\,pc. Thus, the range of possible dust temperatures inferred from our model fits and their uncertainties (650--1450\,K) requires a peak outburst luminosity of (1--25) $\times10^{11}$\,L$_{\odot}$, which is three orders of magnitude higher than the observed peak of the 2014 outburst. Furthermore, the temperature evolution of a thermally reprocessed echo is expected to evolve with time as $T\propto t^{-0.5}$ (Fox et al. 2011, 2015), and so we would have expected the temperature between 133 days and 325 days to have dropped from $\sim830$\,K to $\sim60$\,K; instead, the temperature evolution is consistent with no change between these epochs. We therefore conclude that a light echo is inconsistent with the available data, and therefore is not the source of the late-time UV--optical source and its thermal counterpart. The hypotheses of dust synthesis in the ejecta and swept-up CSM dust are far more consistent with the data.

\subsubsection{Circumstellar extinction and intrinsic stellar parameters}
If the UV--optical component of the SED is from a surviving star and the thermal emission is from circumstellar dust that absorbs stellar radiation, then we should consider the potential effect of dust absorption on the optical properties of the remnant. Under the assumption of a spherically symmetric shell geometry of thickness $\Delta r$, the optical depth of the dust at a given wavelength can be expressed by 
\begin{equation}
\tau_\lambda = \kappa_\lambda(a)\,\rho \, \Delta r = \kappa_\lambda(a) \frac{M_{\rm d}}{4 \pi r_{\rm bb}^2}, 
\end{equation}

\begin{figure}
\includegraphics[width=3.4in]{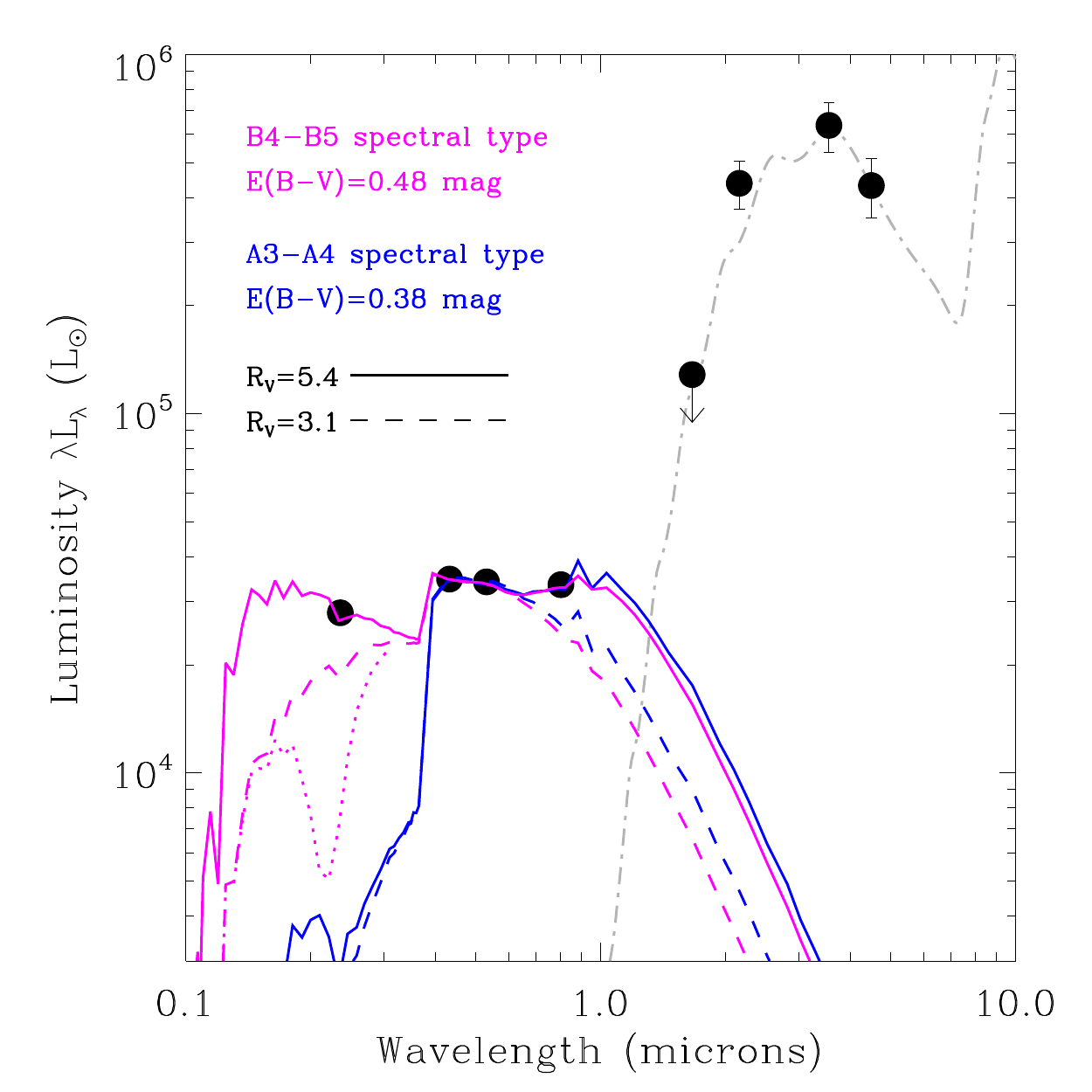}
\caption{SED of the SN\,Hunt248 optical (infrared) remnant at 370 (325) days (black filled circles; corrected only for interstellar extinction $A_V=0.14$\,mag and $R_V=3.1$). The magenta curves are the SED of a model $T_{\rm eff}=15,000$\,K star (spectral type B4--B5; Castelli \& Kurucz 2003), reddened by $E(B-V)=0.48$\,mag (to illustrate the effect of circumstellar extinction) for two different extinction laws. The solid (dashed) curves represent extinction laws having $R_V=5.4 (3.1)$. The reddened models have been vertically scaled to match the $B$ and $V$ photometry. The blue curves are for a model SED of a $T_{\rm eff}=8500$\,K star (spectral type A3--A4) reddened by $E(B-V)=0.38$\,mag, shown to demonstrate that cooler models greatly underestimate the UV photometry. For reference, the dotted magenta curve near 0.2\,${\mu}$m wavelength shows the effect that a Galactic interstellar UV opacity bump would have on the B4--B5, $E(B-V)=0.48$\,mag, $R_V=3.1$ model.  Our silicate IR emission model for $a=1.0$\,$\mu$m is also shown (dashed-dotted grey curve).} 
\label{fig:red_sed}
\end{figure}

\noindent  where $\rho$ is the density of the dust shell and $\kappa_\lambda(a)$ is the absorption coefficient for the dust of a particular grain radius and at a particular wavelength. The $V$-band ($\lambda=0.555\,\mu$m) absorption coefficients for our best-matching graphite ($a=0.3\,\mu$m) and silicate ($a=0.75$--1.0\,$\mu$m) models are $\sim14700$\,cm$^{-2}$\,g$^{-1}$ and $\sim2000$--2600\,cm$^{-2}$\,g$^{-1}$, respectively (see Fox et al. 2010, their Figure\,4). Using the model dust masses in Table\,3 and the radius of $3\times10^{15}$\,cm derived in \S3.1, we estimate $V$-band optical depths of $\tau_V \approx 0.9$ for graphite and $\sim0.7$--0.9 for silicates. If we ignore the effect of grain albedo and optical scattering for the moment, then the extinction can be approximated by 1.086\,$\tau$, in which case we obtain $A_V\approx0.8$--1\,mag.  The total extinction from the ISM and hot-dust component would therefore be approximately the same for both the graphite and silicate models, with $A_V\approx1.0$\,mag. This implies $M_V\approx-7.6$\,mag for the remnant and, thus, supergiant luminosity class. However, a more realistic treatment of the effective extinction accounts for the scattering albedo, $\omega$, of the grains: $A=1.086\,(1-\omega)^{1/2}\,\tau_V$. For a standard ISM-like distribution of graphitic (silicate) grains, the scattering albedo is 0.5\,(0.9) and would thus reduce the required extinction to 0.8\,(0.4) mag (Kochanek et al. 2012a). However, for large grains with $a=0.3$--1.0\,$\mu$m, the effective albedo could be considerably lower ($\omega\approx0.1$; Mulders et al. 2013) and thus scattering might have only a small impact on the effective extinction. Without reliable information on the albedo of the grains, all we can say is that the \textit{expected} extinction from the same hot-dust component that is responsible for the IR emission is $A_V<1$\,mag. 

\begin{figure}
\includegraphics[width=3.3in]{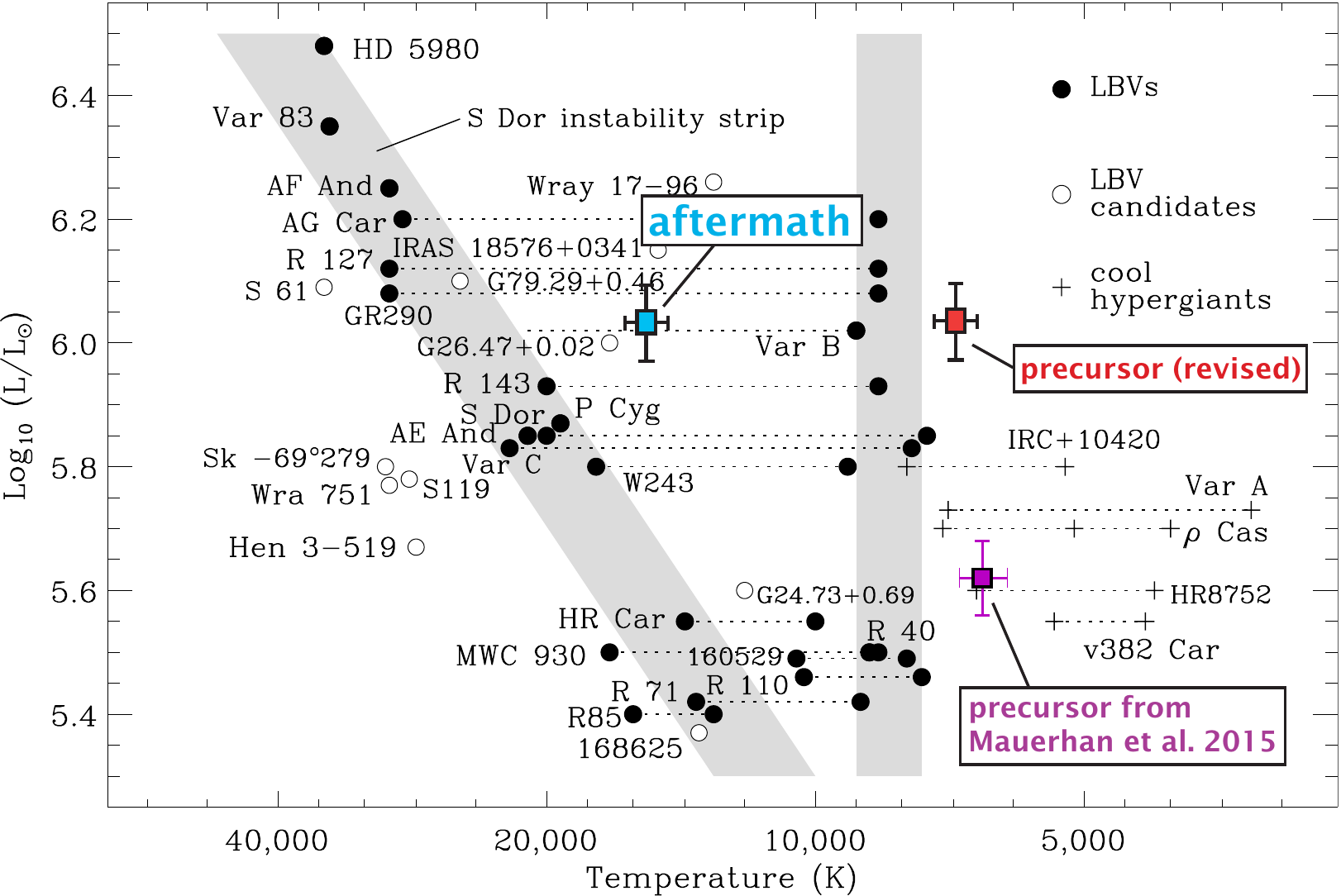}
\caption{Modification of the HR diagram for LBVs and their kin from Mauerhan et al. (2015, see their Figure~8). The magenta coloured square is the previous estimate for the precursor star from Mauerhan et al. (2015), uncorrected for possible circumstellar extinction. The red square represents the revised best-matching model of the precursor SED, corrected for grey circumstellar extinction (see text). The blue square indicates the aftermath of the eruption, a hot B4--B5 supergiant, also corrected for grey circumstellar extinction. The luminosities of the remnant and the revised precursor were calculated by integrating the UV--IR SEDs of the best-matching models shown in Figures~\ref{fig:pre_sed} and \ref{fig:red_sed}. }
\label{fig:HR}
\end{figure}

The effect of the estimated extinction on the colours of the star depends on the assumed value of total-to-selective extinction $R_V$, defined as $A_V/E(B-V)$, which is sensitive to the dust chemistry and grain-size distribution. If we were to hypothetically assume $A_V=1$\,mag and an ISM-like value of $R_V=3.1$ for SN\,Hunt248, then the associated $E(B-V)\approx0.3$\,mag would imply an intrinsic colour in the range $(B-V)_0 = 0.1$ mag, corresponding to a spectral type in the range A3--A4 (Fitzgerald 1970). However, such a spectral type provides a poor match to the UV--optical SED, as illustrated by Figure\,\ref{fig:red_sed}. A3--A4 stars exhibit a relative UV luminosity that is an order of magnitude lower than that of the optical bands. On the contrary, the strong UV flux of the data indicates that the star is significantly hotter, with a substantial Balmer continuum flux. Specifically, after matching stellar model SEDs having a wide range of temperatures (from Castelli \& Kurucz 2003) and over a wide range of $A_V$ and $R_V$, we found that the best match to the four bands of our measured UV--optical SED is provided by a star with $T_{\rm eff}=15,000$\,K (appropriate for a B4--B5 star of supergiant luminosity class; Zorec et al. 2009) with extinction parameters $A_V=2.6$\,mag and $R_V=5.4$ (with no ISM-like UV ``bump" in the extinction law). Cooler models cannot supply enough Balmer continuum, while hotter stellar SEDs with $T_{\rm eff} >15,000$\,K produce too much UV flux, and cannot provide a good match for any of the wide range of $A_V$ and $R_V$ values we attempted. We conservatively estimate a temperature uncertainty of $\delta T=1000$\,K for the remnant star.

Clearly, the extinction value of $A_V=2.6$\,mag implied by our best-matching stellar SED is significantly higher than our estimates of the expected absorption from the hot-dust component, which suggested $A_V<1$\,mag. However, a higher value of extinction would not be surprising, given that the hot dust responsible for the IR emission probably comprises only a fraction of the total dust mass; indeed, it is plausible that there is cooler dust in the system that does not emit strongly at 3--5\,$\mu$m. Furthermore, the value of $A_V=2.6$\,mag implied by the SED would also explain the factor of $\sim10$ drop in apparent brightness of the remnant star relative to the precursor. Meanwhile, the high $R_V$ we inferred from the SED might actually be appropriate for circumstellar dust having a grain distribution skewed toward large sizes. For example, $\sim40$\% of the extinction in the interacting SN\,2010jl has been attributed to large graphitic dust grains with maximum sizes above $a=0.5\,\mu$m and possibly as large as $a>1.3\,\mu$m, which result in an estimated $R_V=6.4$ (Gall et al. 2014). In another example, observations of the red hypergiant VY\,CMa necessitate a circumstellar total-to-selective extinction value of $R_V=4.2$ (Massey et al. 2005), also potentially the result of a grain distribution skewed toward larger sizes. In addition, large grains in $\eta$\,Car's Homunculus nebula have been invoked to explain the apparently grey extinction of the central source (Andriesse, Donn, \& Viotti 1978; Robinson et al. 1987; Davidson et al. 1999; Smith \& Ferland 2007; Kashi \& Soker 2008). 

The integrated extinction-corrected luminosity of the best-matching B4--B5 SED is $L\approx1.2\times10^6\,{\rm L}_{\odot}$. We note that this is approximately twice the luminosity of our previous estimate for the cool hypergiant precursor (Mauerhan et al. 2015). However, that earlier work focused on matching stellar models to the $B$ and $V$ photometry alone (ignoring the poor fit to the $I$-band photometry) and assumed no circumstellar extinction. We thus revisited the precursor photometry in this work using stellar SED models from Castelli \& Kurucz (2003), reddened by an additional component of circumstellar extincition. We find that the best-matching stellar model is one in which there is substantial \textit{grey} circumstellar extinction, similar to our conclusion for the hotter B4--B5 remnant.  As shown in Figure\,\ref{fig:pre_sed}, we obtain a reasonable match to the $B$, $V$, and $I$ precursor photometry using a stellar model SED with $T_{\rm eff}=7000$\,K (Castelli \& Kurucz 2003), reddened by $E(B-V)=0.16$\,mag and $R_V=5.4$ ($A_V=0.86$\,mag). This temperature is more or less equivalent to our previous estimate in Mauerhan et al. (2015), and thus remains consistent with the yellow (F-type) hypergiant classification. We note that if we had used a standard ISM-like $R_V=3.1$, then we achieve poor matches for a wide range of stellar models and extinction values. Moreover, stellar models with lower effective temperatures than 7000\,K exhibit $B$-band fluxes well below the photometry. Based on our attempted matches to models with a variety of temperatures, we conservatively estimate a temperature uncertainty of $\delta T=1000$\,K for the precursor. The integrated unreddened luminosity of our best-matching stellar model (shown in Figure\,\ref{fig:pre_sed}) is also $L\approx1.2\times10^6\,{\rm L_{\odot}}$, equivalent to that of the best-matching B4--B5 remnant model shown in Figure\,\ref{fig:red_sed}. Taken at face value, this is consistent with a temperature change of $\delta T\approx8000$\,K at constant luminosity of $\sim1.2\times10^6\,{\rm L}_{\odot}$. 

The revised luminosity estimate of the precursor warrants an examination of the star's associated transition in the HR diagram, which we show in Figure\,\ref{fig:HR}. After correcting the precursor photometry for the circumstellar extinction discussed above ($A_V=0.86$\,mag and $R_V=5.4$), the precursor star would occupy a region more luminous than the cool hypergiants, yet still within the observed temperature range exhibited by stars of this class (note, however, that circumstellar extinction may not be adequately addressed in other objects classified as cool hypergiants). After the eruption, the hotter remnant has migrated blueward, and lies in between the S\,Dor and red instability strips. Future observations will determine whether the remnant continues to migrate in the HR diagram toward the hotter S\,Dor instability strip occupied by quiescent LBVs, or if increasing extinction from ongoing dust condensation in the ejecta pushes it redward again.

\begin{figure}
\includegraphics[width=3.3in]{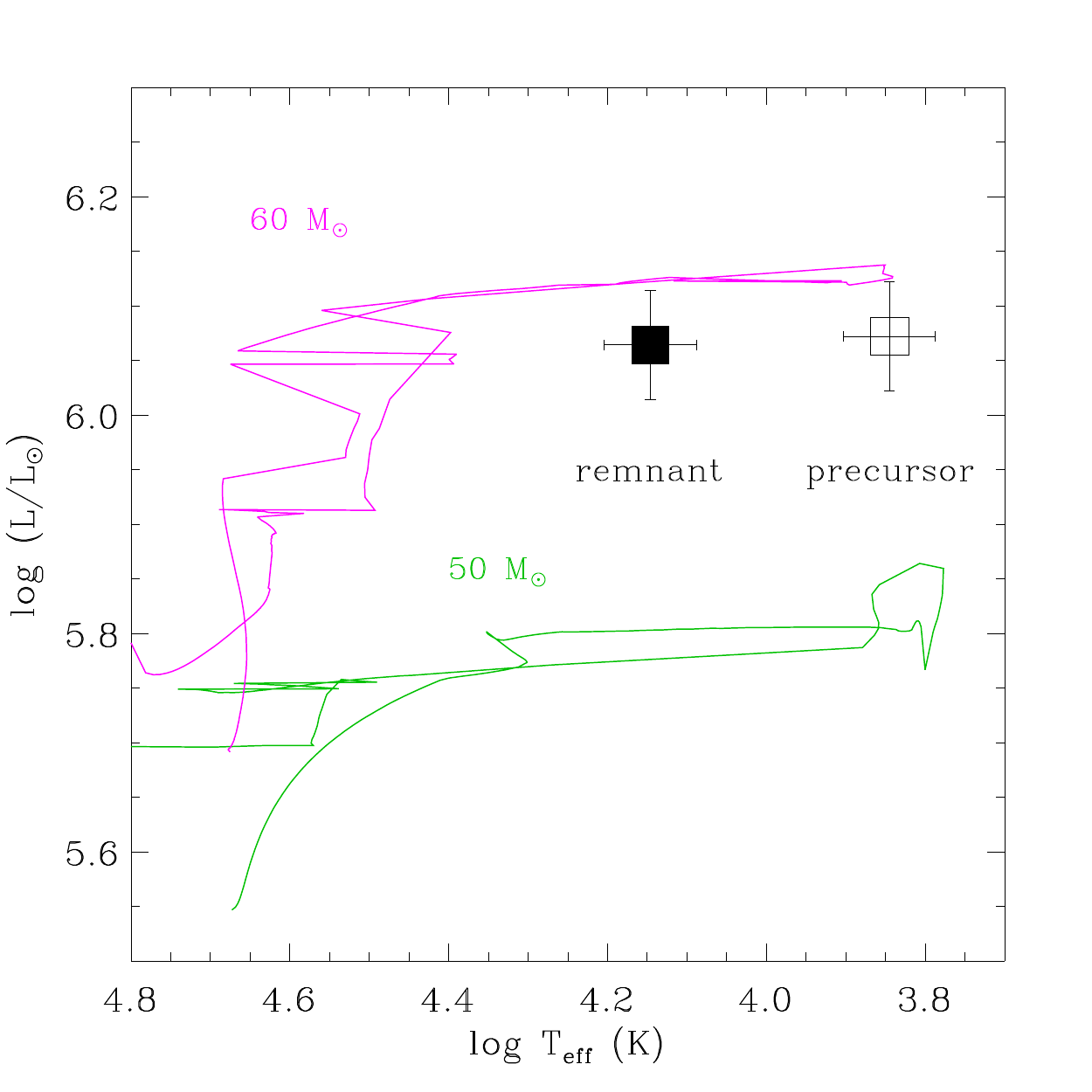}
\caption{The precursor (open black square) and remnant (filled black square) photometry of SN Hunt 248 on the HR diagram, after correcting for our estimated circumstellar extinction parameters (see text \S4.1.2). The evolutionary tracks are from Geneva rotating stellar models (Ekstr{\"o}m et al. 2012) at solar metallicity for initial masses of 50\,M$_{\odot}$ (green curve) and 60\,M$_{\odot}$ (magenta curve). }
\label{fig:pre_iso}
\end{figure}

The revised parameters of the stellar precursor warrant reanalysis of the star's initial mass as well, which was previously estimated at $\sim30\,{\rm M}_{\odot}$ (Mauerhan et al. 2015). Figure\,\ref{fig:pre_iso} shows the data, after correcting for the purported grey circumstellar extinction parameters discussed above (for the remnant, $E(B-V)=0.48$\,mag and $R_V=5.4$; for the precursor, $E(B-V)=0.16$\,mag and $R_V=5.4$), along with evolutionary tracks from the Geneva rotating stellar models for 50\,M$_{\odot}$ and 60\,M$_{\odot}$ (Ekstr{\"o}m et al. 2012); the data appear to most closely match (but are slightly below) the 60\,M$_{\odot}$ model, which at the locations of both the remnant and precursor is undergoing core-He burning (Ekstr{\"o}m et al. 2012). This revised initial mass is approximately twice as high as the circumstellar extinction-free estimate by Mauerhan et al. (2015). Interestingly, the photometry of the remnant and precursor, which was corrected for different values of circumstellar extinction and based on the best SED matches, is consistent with no significant change in stellar luminosity between before and after the 2014 eruption.

\subsection{The 2014 outburst, revisited}
\subsubsection{Massive binary merger-burst?}
The energy source of the 2014 eruption is uncertain. The structure of the outburst light curve, the outflow velocity, and the large-amplitude pre-outburst variability detected over prior decades might provide clues. We can speculate that the cool-hypergiant precursor was a massive interacting binary, perhaps similar to HR\,5171 (Chesneau et al. 2014a), and that its large pseudophotosphere was the signature of a common envelope. If this is the case, it is plausible that SN\,Hunt\,248's 2014 eruption was driven by a violent binary encounter, common-envelope ejection, or a merger-burst marking the coalescence of two massive stars (see Paczy{\'n}ski 1971; Vanbeveren et al. 1998, 2013; Podsiadlowski et al. 2010; Langer 2012; Justham et al. 2014; Portegies-Zwart \& van den Heuvel 2016). Indeed, such events might be more common than previously thought (Kochanek 2014), and their transients might actually explain a substantial fraction of SN impostors and LBV eruptions, including the historic outburst of $\eta$\,Car (Portegies Zwart \& van den Heuval 2016).

\begin{figure}
\includegraphics[width=3.45in]{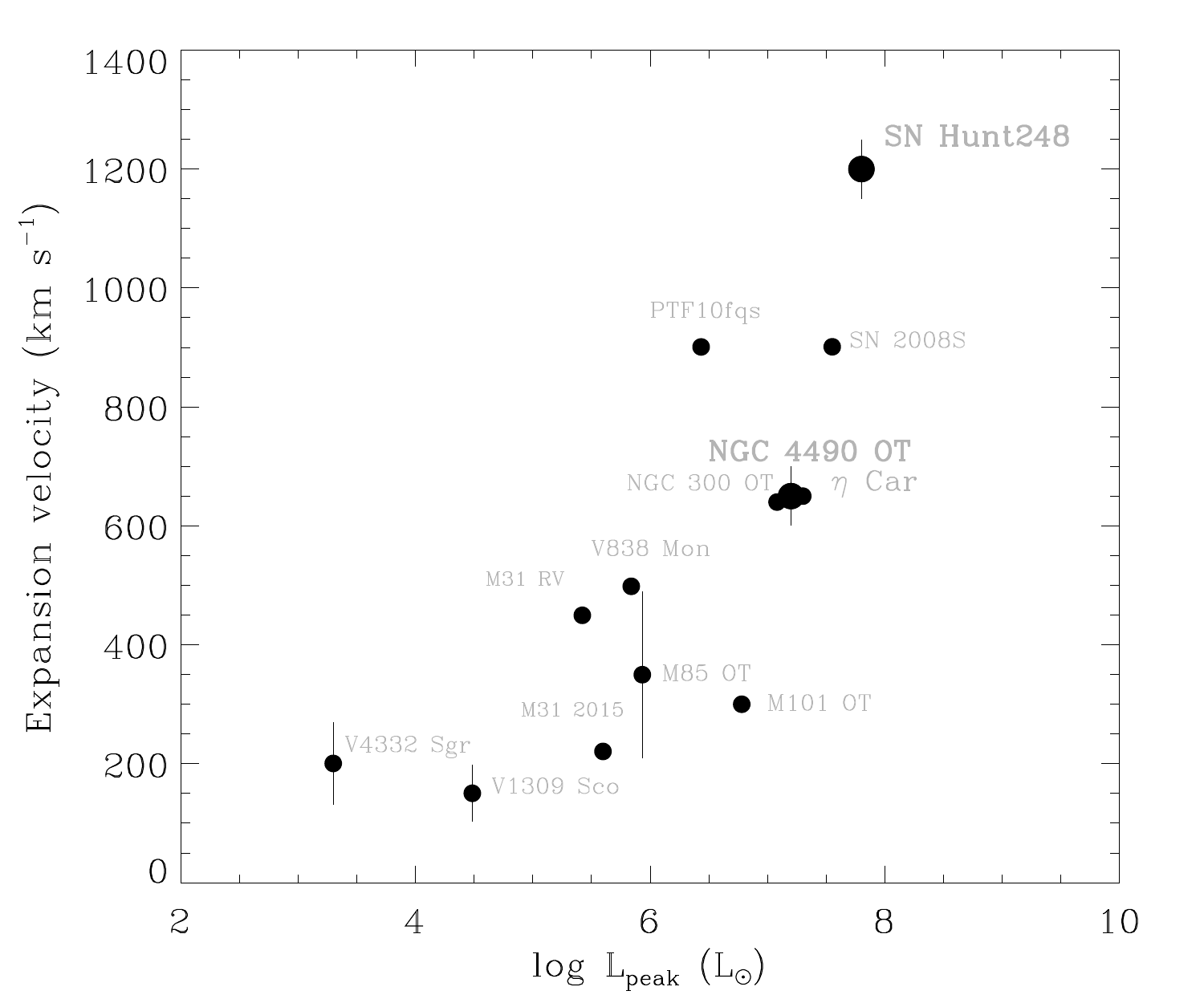}
\caption{Peak outburst luminosity versus outflow velocity for SN\,Hunt\,248 and other stellar merger candidates, including  
NGC\,4490-OT (Smith et al. 2015) and the sample presented by Pejcha et al. (2016a, their Figure~21); $\eta$\,Car is also included. Uncertainties in expansion velocity are shown where available.}
\label{fig:merger}
\end{figure}

Figure\,\ref{fig:merger} shows the peak outburst luminosity versus outflow velocity\footnote{We note that the outflow velocities of SN\,Hunt\,248 and NGC4490-OT were measured from their P\,Cygni absorption minima (Mauerhan et al. 2015; Smith et al. 2016b), whereas the outflow velocities of the sample in Pejcha et al. (2016a) were measured mostly by H$\alpha$ line widths, and $\eta$\,Car's velocity measurement is from detailed spectroscopic analysis of the Homunculus nebula (Smith et al. 2003).} for SN\,Hunt\,248 (Mauerhan et al. 2015); NGC\,4490-OT (Smith et al. 2016b); the sample of merger candidates presented by Pejcha et al. (2016a, their Figure~21); M101-OT, also considered a merger candidate or binary common-envelope ejection event (Blagorodnova et al. 2017); and $\eta$\,Car (Smith et al. 2003; Smith \& Frew 2011). Interestingly, SN\,Hunt\,248 is consistent with the apparent trend exhibited by this sample of merger candidates. Smith et al. (2016b) interpreted NGC\,4490-OT as a stellar merger involving a star of similarly high mass to SN\,Hunt248 ($\sim30$\,M$_{\odot}$), and reiterated the suggestion that $\eta$\,Car's historic eruption was the result of a massive merger. Indeed, $\eta$\,Car's position in Figure\,\ref{fig:merger} also fits in with the apparent trend exhibited by other merger candidates. V1309\,Sco was almost certainly a true merger, based on the exquisite light curve that showed the rapidly decreasing orbital period of an inspiraling binary (Tylenda et al. 2011).  V838 Mon was thought to be a similar merger involving a B-type star (Tylenda et al. 2005; Munari et al. 2007), perhaps in a triple system with another tertiary B-type star (Chesneau et al. 2014b). Both V1309\,Sco and V838\,Mon exhibited double-peaked light curves, of shorter duration and brightness than those of SN\,Hunt\,248 and NGC\,4490-OT, but similar in multipeaked morphology. If they are mergers, the relatively long durations of SN\,Hunt248 and NGC\,4490-OT, compared with V1309\,Sco and V838\,Mon, are to be expected from their relatively high progenitor masses. However, it is important to note that simulations of common-envelope outflows that are shock-energised by the binary's orbital energy input (e.g., Pejcha et al. 2016a; MacLeod et al. 2017) have not yet reproduced the high outflow velocities we have measured for SN\,Hunt248 ($\sim1200$\,km\,s$^{-1}$; Mauerhan et al. 2015), so the apparent trend in Figure\,\ref{fig:merger} has yet to be theoretically established at the high-mass end. More explosive forms of energy input that might result in fast $\sim1000$\,km\,s$^{-1}$ outflow velocities have been proposed to occur during the common-envelope evolution of massive stars (see Podsiadlowski et al. 2010; Soker \& Kashi 2013; Tsebrenko \& Soker 2013), but it is not clear if such effects would result in a continuation or deviation from the apparent trend in Figure\,\ref{fig:merger}. 

A multipeaked light curve might be a natural consequence of a stellar merger or common-envelope ejection. A close binary of evolved massive stars that are headed for a merger will experience  mass transfer, and this can occur even if the primary radius does not fully fill its Roche lobe, but fills it up with material from a slow wind (e.g., wind Roche-lobe overflow, WRLOF; Abate et al. 2013). RLOF may be nonconservative and WRLOF is nonconservative by nature (i.e., some mass is lost rather than exchanged). The process leads to the buildup of CSM with enhanced density in the equatorial plane of the binary, forming a spiral pattern that tightens with increasing radius and forms a dense torus-like structure surrounding the binary (Pejcha et al. 2016a, 2016b; Ohlmann et al. 2016). The subsequent explosive outflow from a merger-burst will encounter this toroidal CSM distribution and generate radiation from the resulting interaction (multiple peaks in the light curve). Any interaction-induced dust formation will mirror the geometry of the pre-existing CSM. Relatedly, the circumstellar environment of the purported post-merger system V838\,Mon exhibits an equatorial overdensity of dust several hundred AU in extent (Chesneau et al. 2014b).  Hydrodynamic simulations have also shown that equatorially enhanced dust formation should be expected in the aftermath of mergers (Pejcha et al. 2016a). 

In comparing SN\,Hunt248 to stellar mergers, we should note that the B4--B5 spectral type we have estimated for the remnant would be much hotter than that of the immediate aftermath of the purported \textit{complete} merger V838 Mon, which became the coolest supergiant ever observed with $T\approx2000$\,K (L3 spectral type; Loebman et al. 2014). The cool source is presumably the inflated merger product, apparently contracting on a thermal timescale (Chesneau et al. 2014b). The relatively hot spectral type of the remnant of SN\,Hunt248 suggests that, if the eruption did indeed stem from a merging binary, then the individual stars might have avoided a complete merger while ejecting their common envelope. We speculate that the purported pseudophotosphere of the cool hypergiant was destroyed with the ejection of the inflated common envelope, revealing the stellar photosphere(s) of the hotter B4--B5 star(s) inside. 

With regard to binary origin, it is possible that the light of the remnant is dominated by a companion to the eruptive source, or perhaps even a third tertiary companion to a progenitor binary system that may have merged. The latter idea, although very speculative at this point, is motivated by the discovery of a tertiary B-type companion in the V838 Mon system, which eventually became heavily reddened by expanding ejecta dust $\sim5$\,yr after the event (Wisniewski et al. 2008; Tylenda et al. 2011). This comparison warrants continued monitoring of SN\,Hunt248 and NGC\,4490-OT. Tertiary stars of triple systems could play an important role in the merger of the tighter pair, as has been suggested for $\eta$\,Car (Portegies Zwart \& van den Heuval 2016). 

\subsubsection{Peculiar core-collapse supernova?}
Finally, we discuss the possibility that the 2014 eruption of SN\,Hunt248 was a terminal explosion. This speculation is warranted, given renewed deliberation on the fates of transients previously classified as nonterminal SN impostors, including the prototype SN impostor SN\,1997bs, and SN\,2008S (Adams \& Kochanek 2015; Adams et al. 2016). Like SN\,Hunt248, SN\,1997bs exhibited relatively narrow spectral lines (no obvious sign of high-velocity ejecta), peaked at a luminosity below that of typical core-collapse SNe, and had a much shorter duration than common SNe~II-P. Interestingly, at $\sim1$\,yr post-eruption, SN\,1997bs exhibited an optical remnant very similar in brightness to that of SN\,Hunt248 (see Figure\,\ref{fig:lc}), but which continued to fade during subsequent coverage (Kochanek et al. 2012b). Remarkably, the most recent optical-IR data on SN\,1997bs appear to be consistent with a terminal explosion, as few plausible combinations of obscuring dust and surviving stellar luminosity can explain the late-time data. Could SN\,Hunt\,248 have been a terminal event, similar to what has been suggested for SN\,1997bs? If so, then the UV--optical remnant could either be a companion star, or it could be residual SN emission that coincidentally appears similar to the attenuated B4--B5 supergiant SED we constructed. In the latter possibility, we might expect the light curve of the optical remnant to continue evolving similarly to SN\,1997bs (see Figure\,\ref{fig:lc}), underscoring the need for continued UV--IR observations.

\section{Summary and concluding remarks}
We have presented space-based observations of the aftermath of SN\,Hunt248 with \textit{HST} and \textit{Spitzer}. The UV--optical SED is consistent with a B4--B5 supergiant attenuated by grey circumstellar extinction. Our modeling of the \textit{Spitzer} data suggests that the dust responsible for the IR emission is composed of relatively large grains ($a \gtrsim 0.3\,\mu$m), has a mass of $\sim10^{-6}$--$10^{-5}\,{\rm M}_{\odot}$ (depending on whether it is graphitic or silicate), and a temperature of $T_d\approx900$\,K. The large grain size indicated by our modeling results is consistent with the grey extinction we infer for the UV--optical remnant. However, the extinction expected from the hot-dust component alone is significantly below the amount suggested by the best-matching UV--optical SED, prompting us to speculate on the presence of cooler dust not detected by our 3.6 and 4.5\,$\mu$m photometry. Future mid-IR observations with the \textit{James Webb Space Telescope}~(\textit{JWST}) could reveal such cooler dust.

We revised our analysis of the precursor-star photometry, and showed that the SED is well matched by an F-type supergiant that also suffers grey circumstellar extinction but of lesser magnitude than the remnant. Comparison of the extinction-corrected photometry to rotating stellar models indicates that the initial mass of the star could be nearly $\sim60\,{\rm M}_{\odot}$, approximately twice the value estimated by Mauerhan et al. (2015).

We interpreted the 2014 outburst of SN\,Hunt248 in the context of binary mergers, as in the very similar case of NGC\,4490-OT (Smith et al. 2016b). If such an interpretation is correct, then the hot B4--B5 spectral type of the byproduct might suggest that the binary avoided a complete merger during the ejection of the common envelope. In this interpretation, it could be that the ejection of the common envelope resulted in the destruction of the cool hypergiant pseudophotosphere suggested by Mauerhan et al. (2015), and prompted the star's transition to B4--B5 spectral type. This hypothesis, of course, requires that the remnant light, particularly the UV flux, is dominated by the eruptive star, and not a binary companion or unrelated neighboring source.

The nature of the stellar aftermath and the 2014 eruption will be elucidated further with future UV through IR monitoring of the source using {\it HST} and \textit{JWST}. Specifically, additional observations to track the evolution of the SED will allow for the construction of more complex models involving a surviving central source(s) attenuated by an evolving dust component. If dust has continued to condense in the ejecta since the last observations, then we expect that the UV--optical extinction will increase, regardless of whether the central light is from the eruptive source or a binary companion that has also been engulfed by the ejecta. If dust formation has ceased, however, then we might observe the future restrengthening of the optical flux, as the optical depth of an expanding dusty ejecta should decrease with geometric expansion over time as $\tau \propto t^{-2}$ (Kochanek et al. 2012b). In both cases, if dust is in a continually expanding unbound outflow, then it will also have cooled, and the 1--5\,$\mu$m IR excess will fade as the flux shifts to longer wavelengths. On the other hand, if future observations reveal that the dust has remained hot and emitting at near-IR wavelengths, it would indicate that there could be an additional circumstellar dust component close to the stellar source, perhaps similar to the case of $\eta$\,Car (e.g., Smith et al. 2010) or dusty Wolf-Rayet binaries (e.g., Williams et al. 2012).

\section*{Acknowledgements}
\scriptsize 
This work is based in part on observations made with the NASA/ESA {\it Hubble Space Telescope}, obtained from the Data Archive at the Space Telescope Science Institute (STScI), which is operated by the Association of Universities for Research in Astronomy (AURA), Inc., under NASA contract NAS5-26555. This work is also  based in part on observations and archival data obtained with the {\it Spitzer Space Telescope}, which is operated by the Jet Propulsion Laboratory, California Institute of Technology, under a contract with NASA; support was provided by NASA through an award issued by JPL/Caltech. A.V.F.'s supernova group is also supported by Gary \& Cynthia Bengier, the Richard \& Rhoda Goldman Fund, the Christopher R. Redlich Fund, the TABASGO Foundation, and the Miller Institute for Basic Research in Science (U.C. Berkeley).

\end{document}